# Computable Fairness:
## Boltzmann–Softmax Control for AI Resource Allocation

**Ji-Won Park[1] and Chae Un Kim[2]**

## Abstract

In large-scale artificial intelligence systems, the allocation of scarce resources—such as GPU compute time, network bandwidth, and training opportunities—among multiple agents has become an increasingly critical challenge. Conventional resource allocation policies, however, are primarily designed around efficiency metrics such as throughput and performance, potentially leading to dominance concentration in which a small number of agents disproportionately occupy system resources. Such concentration can undermine system diversity and long-term stability, with particular risks in distributed learning and federated learning environments where structural imbalance may be amplified.

To address this problem, we propose a framework grounded in the principle of computable fairness, formalized as Computable Fair Division (CFD), for AI resource allocation. The key contribution of this work is not the introduction of a new mathematical formula. The Boltzmann–Softmax function has been widely used in deep learning and reinforcement learning, but its role has been limited to serving as a differentiable approximation of argmax—a tool for selection rather than distribution. We reinterpret this same function as a probabilistic resource allocation mechanism, redefining the inverse temperature parameter β not as a fixed numerical constant but as a computable control variable that dynamically governs the balance between efficiency and fairness. As β increases, resources concentrate toward high-contribution agents, enhancing efficiency; as β decreases, the distribution flattens, improving fairness. Under this formulation, CFD implements fairness not as an externally imposed

[1] Regional Science, Cornell University, Ithaca, NY, 14853, USA; Department of Economics, University of Ulsan, Ulsan, 44610, Korea. E-mail: jp429@cornell.edu

[2] Department of Physics, UNIST, Ulsan, 44919, Korea. E-mail: cukim@unist.ac.kr

Code: https://github.com/entrofy-ai/computable-fairness

normative constraint or post-hoc correction, but as an operational principle that is computed and dynamically adjusted within the system.

The proposed framework consists of two stages: static optimization and dynamic feedback control. Static analysis reveals that the trade-off between efficiency and fairness exhibits a Pareto frontier structure, with a near-optimal operating region—rather than a single optimum—where total loss remains approximately constant. We define this region as the Stability Corridor and use the optimal $\beta^*(\lambda)$ derived from it as the reference point for dynamic control. In the dynamic setting, the AHC++ (Adaptive Hard-Cap Controller++) operates as a closed-loop controller that updates $\beta$ in real time using the error between observed dominance and a policy-specified target as the feedback signal. Simulation results demonstrate that the proposed control structure suppresses transitions to extreme dominance concentration even under exogenous shocks, while tracking the fairness target without substantial throughput degradation. Computational scalability analysis further shows that as the number of agents N increases by a factor of 100, execution time increases by a factor of only approximately 5.5, substantially more moderate than the $O(N^2)$ scaling of pairwise comparison-based approaches. These results indicate that the framework is capable of repeated execution in large-scale multi-agent environments.

In summary, this work reframes AI resource allocation as a joint problem of probabilistic allocation rules and feedback control, presenting a computable fair division mechanism applicable across GPU scheduling, federated learning, and distributed AI systems.

The source code and reproducible simulation environment for the CFD framework (AHC++) are publicly available at: https://github.com/entrofy-ai/computable-fairness

# 1. Introduction

The rapid advancement of generative artificial intelligence, particularly the proliferation of large language models (LLMs) and foundation models, has expanded the role of AI beyond prediction and generation into operational decision-making in complex digital environments (Bommasani et al., 2021; OpenAI, 2023). This shift has elevated a new class of resource allocation problems—involving massive computational resources, limited service capacity, and competition among numerous users, agents, and tasks—as a central challenge. In large-scale AI infrastructure, the key question is no longer confined to "how accurate is the model." How to allocate scarce computational resources—to whom, by what principle, and how fairly—has become a fundamental issue that governs system stability and long-term sustainability.

Nevertheless, many AI systems are still designed from the perspective of single-objective optimization, such as error minimization, reward maximization, or throughput improvement (Sutton & Barto, 2018). While such efficiency-oriented designs can enhance short-term performance, in dynamic environments where repeated feedback operates, they can amplify small advantages into cumulative concentration, reinforcing resource monopolization by specific agents (O'Neil, 2016). The runaway feedback loops reported in the context of predictive policing demonstrate that such structural risks are not merely theoretical concerns but observable phenomena in real systems (Ensign et al., 2018). Fairness in AI resource allocation should therefore be treated not as a matter of post-hoc ethical review, but as an operational design requirement for preventing system collapse and monopolization.

## 1.1 Related Work

**Algorithmic fairness.** Individual fairness established the principle that similar individuals should be treated similarly (Dwork et al., 2012), while statistical and group-level fairness criteria such as disparate impact and equality of opportunity have provided important frameworks for diagnosing and correcting bias in predictive outcomes (Feldman et al., 2015; Hardt et al., 2016; Barocas et al., 2023). However, these works primarily focus on fairness in classification and prediction, and their extension to immediately executable allocation mechanisms in repeated, real-time resource distribution settings remains limited.

**Contribution-based allocation.** The Shapley value offers a theoretically principled concept for evaluating cooperative contributions, but its computational cost is prohibitively high in large-scale systems. Data Shapley exemplifies both the theoretical appeal and the computational burden of this approach (Shapley, 1953; Ghorbani & Zou, 2019). Moreover, it is well established that multiple fairness criteria may be mutually incompatible, and that structural trade-offs between fairness and performance can arise (Kleinberg et al., 2017; Zafar et al., 2017).

**Efficiency–equity trade-offs in economics.** The structure of a unified objective function that balances efficiency and equity through a single parameter has a long tradition in economics. Atkinson's (1970) inequality index provides a framework in which a single parameter $\varepsilon$ continuously interpolates between pure efficiency ($\varepsilon = 0$) and Rawlsian equality ($\varepsilon \to \infty$). Okun (1975) systematized the costs of the efficiency–equity trade-off through the "leaky bucket" argument, which holds that efficiency losses are inevitable in redistribution. The social welfare function in welfare economics shares a similar mathematical structure (Samuelson, 1947; Arrow, 1951). In these approaches, however, the policy parameter has typically been imposed externally as a product of philosophical or political choice, rather than internalized as an operational variable that the system can compute and adjust in real time.

**Boltzmann distribution in allocation problems.** Several prior works have applied the Boltzmann distribution to real-world allocation problems. Park et al. (2012) applied the Boltzmann distribution to the initial allocation of emission permits, and Park and Kim (2021) formalized income distribution as a Boltzmann structure. Subsequently, Park et al. (2022) proposed Boltzmann fair division from the perspective of distributive justice, and Park (2024) extended this approach to the environmental policy context. These studies support the socioeconomic validity of Boltzmann-based allocation principles.

## 1.2 Approach and Contributions

The aforementioned lines of research each provide important insights, yet they share common limitations. Algorithmic fairness research focuses primarily on classification and prediction, making it difficult to apply directly to repeated resource allocation. Contribution-based methods such as the Shapley value incur excessive computational costs in large-scale systems. In economic efficiency–equity frameworks, the balance has been an object of analysis rather than a target of real-time control in dynamic systems. AI resource allocation therefore requires a form of fairness that is not only normatively meaningful but also repeatedly executable and scalable—that is, computable fairness. This work formalizes such computable fairness through the Computable Fair Division (CFD) framework. Here, computable fairness refers to the design principle that this work pursues, and CFD is the concrete framework that implements it.

CFD is an approach that implements fairness not as a normative declaration or post-hoc correction rule, but as an algorithmic allocation principle that can be repeatedly executed in large-scale systems. Accordingly, the central question of this work does not stop at "what allocation is philosophically just." Rather, it asks: "what allocation principle is computable in a dynamic AI system environment, and can be executed in practice without compromising system stability?"

Under this problem formulation, we reinterpret the Boltzmann–Softmax function not as a tool for selection but as a mechanism for distribution, and redefine the inverse temperature parameter $\beta$ as a computable control variable that governs the balance between efficiency and fairness. This reinterpretation is not an arbitrary design choice. It is grounded in the maximum entropy principle, which derives the least biased probability distribution under given constraints without additional assumptions, thereby securing information-theoretic justification (Jaynes, 1957; Cover & Thomas, 2006). Boltzmann fair division provides the allocation principle—a probabilistic distribution rule grounded in entropy maximization. CFD extends this principle into an online computable control framework for AI systems, adding dynamic feedback control (Adaptive Hard-Cap Controller++; AHC++), hard-cap constraints, and the stability corridor as operational design elements that were absent in the original formulation.

Furthermore, this work draws attention to the possibility that efficiency and fairness are not necessarily in a zero-sum relationship. While the existing fairness literature has convincingly demonstrated the incompatibility of multiple fairness criteria and the potential for performance loss (Kleinberg et al., 2017; Zafar et al., 2017), this does not imply that efficiency and fairness invariably collide in dynamic resource allocation environments. Based on a unified objective function and a dynamic control structure, this paper explores whether near-optimal operating regions exist in which performance degradation is limited while excessive concentration is suppressed. We define this region as the Stability Corridor, and thereby reframe the relationship between fairness and efficiency not as a simple dichotomy but as a problem of controllable operational space.

The contributions of this paper are as follows.

**(1) Reinterpretation of Softmax and β as a control variable.** We reinterpret the Boltzmann–Softmax function—mathematically identical in structure to the Boltzmann distribution in statistical physics—as a probabilistic resource allocation mechanism, and redefine the inverse temperature parameter β as a computable control variable that dynamically governs the efficiency–fairness balance.

**(2) AHC++ closed-loop control structure.** We newly propose the Adaptive Hard-Cap Controller++ (AHC++), which updates β in real time using the error between observed dominance and the target value as a feedback signal.

**(3) Identification of the Stability Corridor.** Through static loss landscape analysis, we identify a near-optimal operating region (Stability Corridor) in which both efficiency and fairness can be simultaneously maintained, suggesting the possibility of more stable operation even under policy parameter uncertainty.

**(4) Computational scalability.** We confirm substantially more moderate execution time growth compared to pairwise comparison-based $O(N^2)$ approaches, demonstrating the feasibility of repeated execution in large-scale multi-agent environments.

This work transforms fairness from a normative concept into an executable allocation principle, and presents a theoretical and engineering framework for resource management in large-scale AI systems.

# 2. Methodology

This work reformulates the econophysics model of Park and Kim (2021), which defines an optimal allocation state that maximizes social welfare, as a resource allocation and real-time optimization problem for artificial intelligence (AI) systems. The validity of this model has been independently confirmed in a recent cross-country analysis of feasible income equality using the Boltzmann distribution (Sitthiyot & Holasut, 2025). Whereas the prior work established a static criterion for “feasible equality,” the present study extends it by combining it with dynamic feedback control to construct a computable framework.

The core objective is to implement fairness not through external regulation or post-hoc correction, but as computable design variables embedded within the allocation rule and control structure, dynamically adjusted through the interaction between a Boltzmann-based probabilistic allocation rule and an adaptive controller.

The model involves several variables and control parameters. Because these symbols are used repeatedly throughout the subsequent equations and algorithm descriptions, the definitions of key variables are summarized in Table 1 for the reader's convenience. In particular, the contribution score $z_i$ representing each agent's contribution, the inverse temperature parameter β governing allocation concentration, and the probability $cap(t)$ serve as the core control variables of this model.

**Table 1.** Key variables used in the Boltzmann–Softmax resource allocation model.

| Symbol | Description | Symbol | Description |
|---|---|---|---|
| $i$ | Agent or task index | $cap(t)$ | Hard-cap (upper bound on agent probability) |
| $z_i$ | Contribution or performance score of agent i | $Dom(t)$ | System dominance (max allocation probability) |
| $P_i$ | Resource allocation probability assigned to agent i | $\rho_{target}(t)$ | Policy-specified target dominance |
| $x_i$ | Actual allocated resource amount | $\rho_{eff}(t)$ | Effective target dominance considering active agent count |
| $D_i$ | Resource demand of agent i | $L_{eff}$ | Efficiency loss term |
| $K(t)$ | Number of active agents at time t | $L_{ineq}$ | Fairness loss term |
| $\beta$ | Inverse temperature parameter controlling allocation concentration | $L_{tot}$ | Unified loss function |
| $\beta(t)$ | Time-varying inverse temperature parameter | $\lambda$ | Policy weight between efficiency and fairness |

***Note.*** *N* denotes the total number of agents (population size), used as a fixed constant in static analysis. K(t) ≤ N is the number of active agents requesting resources at time t, varying dynamically over time. The loss terms $L_{eff}$, $L_{ineq}$, $L_{tot}$ are all functions of β; in static analysis they are written explicitly as $L_{eff}(\beta)$, $L_{ineq}(\beta)$, $L_{tot}(\beta,\lambda)$.

Figure 1 presents an overview of the CFD framework. The left panel (static model) takes agent scores as input, computes Boltzmann–Softmax probabilities governed by β, and derives the static optimum via a unified objective function. The region where performance remains robust to policy weight variation defines the stability corridor. The right panel (dynamic model) depicts an online scheduler that allocates resources under time-varying conditions, applies a hard-cap followed by renormalization, and feeds back the dominance error to AHC++, which updates β in a closed loop. The following subsections describe each component in sequence.

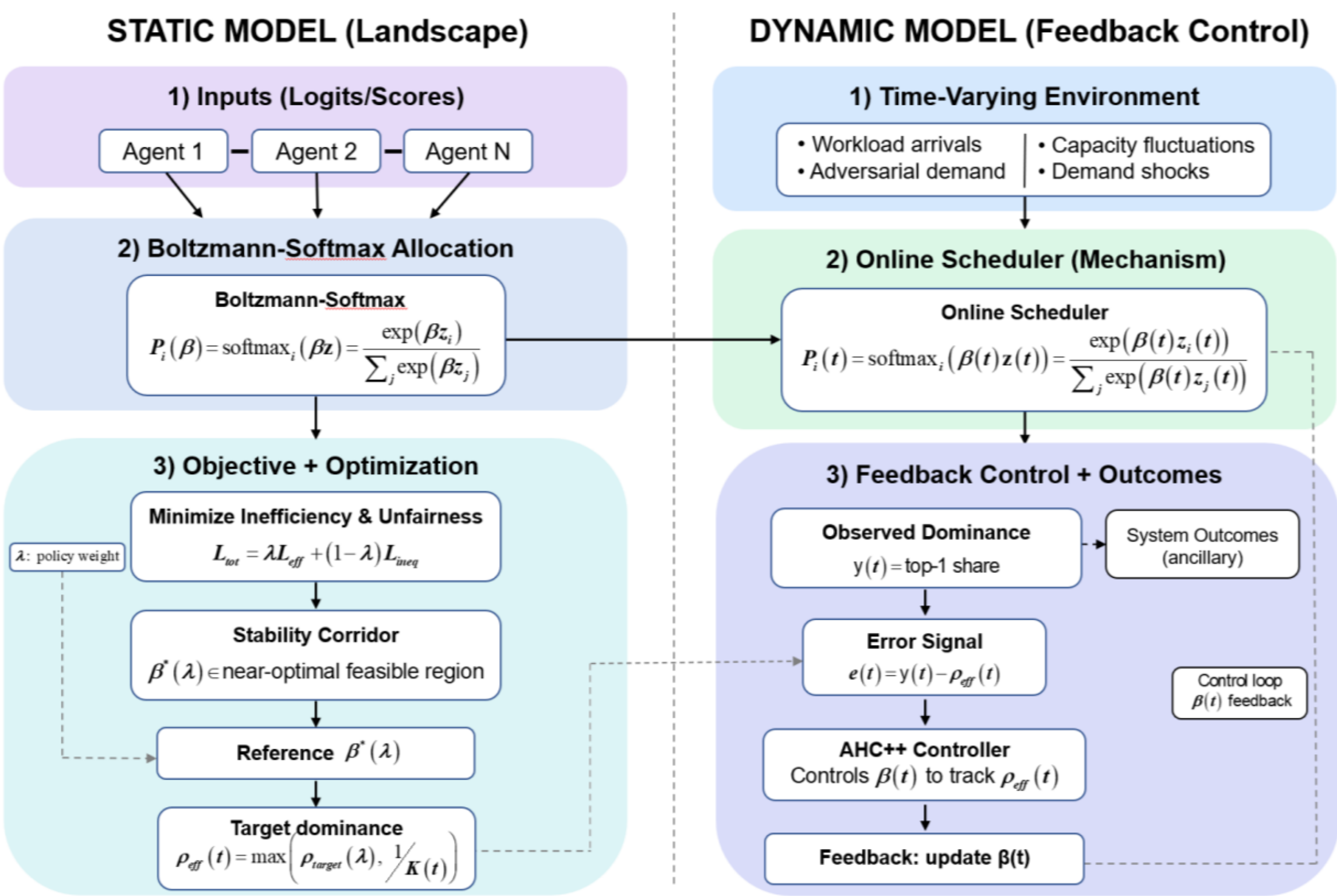

**Figure 1.** CFD framework: Boltzmann–Softmax feedback control architecture. Static model (left) derives optimal β and target dominance. Dynamic model (right) forms a closed-loop with AHC++ updating β in real time.

### 2.1. Variable Mapping: From Economics to AI Engineering

The core elements of traditional income distribution models—individual, income, and capability—are redefined in multi-agent system (MAS) and federated learning environments as follows.

**Table 2.** Variable mapping from economics to AI engineering.

| **Model component** | **Economics** | **AI Engineering** | **Symbol** |
|---|---|---|---|
| Actor | Individual | Agent (node) | $i$ |
| Allocated object (resource) | Income / Wealth | Allocated resource amount | $x_i$ |
| Allocation share (probability) | Income share | Allocation share / probability | $P_i$ |
| Determining factor | Capability / Contribution | Score / Contribution | $z_i$ |
| Control variable | Inequality sensitivity (Inverse temperature) | Exploration–exploitation parameter (Inverse temperature) | $\beta$ |
| Policy variable | Policy weight | Policy weight | $\lambda$ |

The variable mapping in Table 2 extends beyond conceptual analogy to a direct correspondence in real AI systems. For example, the allocation variable $P_i$ (or $x_i$) is realized in GPU scheduling environments as allocated GPU compute time, memory bandwidth, or credit/incentive rewards for model updates. The contribution variable $z_i$, on the other hand, can be measured in federated learning or distributed learning systems as accuracy contribution, per-node throughput, or resource efficiency. These practical correspondences demonstrate that the theoretical structure of this work can be directly applied to resource allocation problems in large-scale AI systems.

## 2.2. Reinterpretation of the Softmax Function and the Boltzmann Allocation Protocol

The softmax function is mathematically identical to the Boltzmann distribution from statistical physics. This work reinterprets it not for selection but as a principle for resource distribution.

When $N$ agents exist in the system, the allocation probability $P_i(t)$ assigned to agent $i$ at time $t$ is determined by the agent's contribution score $z_i(t)$ and the system's inequality sensitivity parameter β as follows:

$$P_i(t) = \text{softmax}_i(\beta \mathbf{z}(t)) = \frac{e^{\beta z_i(t)}}{\sum_{j=1}^{N} e^{\beta z_j(t)}} \tag{1}$$

From a physics perspective, as the temperature $T$ decreases (i.e., as the inverse temperature β increases), the probability of lower-energy states increases sharply, giving rise to a tendency toward monopolization by a specific state. Conversely, as $T$ increases, the probability distribution becomes more uniform.

From an AI engineering perspective, agents with higher contribution scores $z_i(t)$ receive a larger share of resources, but as β decreases, even agents with low contributions obtain a certain level of resources, increasing exploration. This serves to suppress winner-take-all phenomena and preserve system diversity.

The parameter β thus determines the allocation philosophy of the system, extending the softmax-based rule beyond selection into a fair division framework governing the efficiency–fairness balance in a computable manner.

- **β → ∞ (low temperature):** In the low-temperature regime where β is very large, the agent with the highest contribution score nearly monopolizes resources, approaching a Greedy/Selection mode.
- **β → 0 (high temperature):** Resources are distributed uniformly regardless of contribution, converging to a Uniform/Random mode.
- At intermediate levels of β, a region emerges in which an incentive structure proportional to contribution is maintained while the probability distribution does not become excessively concentrated. This allocation regime was conceptualized as Boltzmann Fair Division in prior work (Park et al., 2022), and the present study extends it into a computationally implementable structure, interpreting it within the Computable Fair Division framework.

Computable Fair Division operationalizes this allocation regime through probabilistic rules and control variables that internalize fairness as a computable function of system state, rather than imposing it through external rules or post-hoc corrections.

### 2.3. Unified Loss Function and Optimization Formulation under Hard-Cap Constraints

Conventional AI training has primarily focused on minimizing or maximizing a single efficiency criterion, such as prediction error or reward. In contrast, this work defines a unified loss function (unified objective function) that enables simultaneous control of efficiency and fairness within a single objective, applied to the probability distribution $P(\beta)=\{P_i(\beta)\}_{i=1}^{N}$ generated by the Boltzmann–Softmax allocation mechanism:

$$L_{tot}(\beta,\lambda)=\lambda L_{eff}(\beta)+(1-\lambda)L_{ineq}(\beta),\quad \lambda\in[0,1] \tag{2}$$

Here, β is the control variable governing allocation concentration (inverse temperature), and λ is the policy weight set by the system operator. The meaning of each term is as follows.

The efficiency loss $L_{eff}(\beta)$ is a term that increases when resources are not sufficiently allocated to high-contribution agents. In this work, it is defined as a weighted sum of normalized contribution scores and allocation probabilities. First, the contribution scores are normalized to the [0, 1] interval:

$$\overline{z}_i=\frac{z_i-z_{\min}}{z_{\max}-z_{\min}}$$

Using this, the efficiency loss is defined as follows:

$$L_{eff}(\beta)=1-\sum_{i=1}^{N}P_i(\beta)\cdot\overline{z}_i \tag{2a}$$

Under this definition, $L_{eff}(\beta)$ decreases as more resources are allocated to high-contribution agents, and increases when resources are distributed uniformly regardless of contribution.

The fairness loss term, on the other hand, is defined as an entropy-based measure of allocation probability concentration. The entropy of the probability distribution $P(\beta)$ is expressed as:

$$H\left(P(\beta)\right) = -\sum_{i=1}^{N} P_i(\beta)\log P_i(\beta) \tag{3}$$

Accordingly, the fairness loss is expressed as:

$$L_{ineq}(\beta) = -H\left(P(\beta)\right) = \sum_{i=1}^{N} P_i(\beta)\log P_i(\beta) \tag{4}$$

That is, minimizing $-H\left(P(\beta)\right)$ is equivalent to maximizing $H\left(P(\beta)\right)$, and this operates in the direction of mitigating the winner-take-all phenomenon in which probability mass is excessively concentrated on a specific agent.

$L_{ineq}(\beta)$ thus measures allocation concentration; minimizing it is equivalent to increasing allocation diversity.

The policy weight $\lambda \in [0,1]$ is a policy parameter chosen by the operator (or policymaker). As λ increases, the weight of the efficiency term grows, prioritizing performance; as λ decreases, the weight of the fairness term grows, emphasizing allocation diversity and monopoly suppression. Therefore, λ can be interpreted as a single policy variable that continuously adjusts the balance point between efficiency and fairness.

The static optimization problem in this work is thus formulated as finding the allocation concentration parameter β that minimizes the unified loss function $L_{tot}(\beta,\lambda)$ under a given policy weight λ.

**2.4. Optimization Problem Definition with Hard-Cap Constraints**

The objective of this work is to derive the optimal control parameter $\beta^*(\lambda)$ that minimizes the unified loss under a given policy weight λ, while simultaneously satisfying a hard-cap constraint that prevents structural monopolization by any specific agent. In the dynamic setting, the target dominance $\rho_{target}(t)$ and the number of active agents K(t) are used to define the effective target dominance (upper bound) $\rho_{eff}(t)$, which guarantees achievability, as follows:

$$\rho_{eff}(t) = \max\left(\rho_{target}(t), \frac{1}{K(t)}\right) \tag{5}$$

Here, $\frac{1}{K(t)}$ represents the minimum probability when active agents receive resources equally, thereby preventing the dominance upper bound from being set to a physically unachievable value. In the static analysis (static landscape/corridor analysis), a representative upper bound $\bar{\rho}$ is used in place of the time-varying $\rho_{eff}$, and the following constrained optimization problem is considered:

$$\beta^*(\lambda) \in \arg\min_{\beta} L_{tot}(\beta;\lambda) \tag{6a}$$

subject to

$$\max_i P_i(\beta) \le \bar{\rho} \quad \text{(Hard-Cap Constraint)} \tag{6b}$$

where $P_i(\beta)$ is defined by the Boltzmann–Softmax distribution:

$$P_i(\beta) = \text{softmax}_i(\beta\mathbf{z}) = \frac{e^{\beta z_i}}{\sum_{j=1}^{N} e^{\beta z_j}} \tag{7}$$

The above optimization problem is designed to minimize the unified loss function $L_{tot}(\beta,\lambda)$ determined by the policy weight λ, while simultaneously preventing excessive concentration of probability mass on any specific agent.

That is, by combining the objective function that jointly considers efficiency and fairness with the hard-cap constraint, the system is naturally guided to operate within the efficiency–fairness balance region (stability corridor).

In the dynamic environment, the primary probability distribution is first computed with respect to the time-varying scores $z_i(t)$ and the control variable $\beta(t)$:

$$P_i'(t) = \text{softmax}_i(\beta(t)\mathbf{z}(t)) = \frac{e^{\beta(t)z_i(t)}}{\sum_{j=1}^{K(t)} e^{\beta(t)z_j(t)}} \tag{8}$$

A hard-cap constraint is then applied to ensure that each agent's probability does not exceed the dominance upper bound:

$$\tilde{P}_i(t) = \min\left(P_i'(t), \rho_{eff}(t)\right) \tag{9}$$

Finally, renormalization is performed so that the total probability sums to one, yielding the final resource allocation probabilities:

$$P_i(t) = \frac{\tilde{P}_i(t)}{\sum_{j=1}^{K(t)} \tilde{P}_j(t)} \tag{10}$$

This procedure combines a hard-cap constraint that directly limits the concentration of the probability distribution with renormalization for probability mass conservation, thereby structurally preventing the system from excessively concentrating resources on any specific agent.

Specifically, renormalization redistributes the excess probability mass only among non-capped agents; if any agent exceeds the cap after redistribution, the clipping-and-redistribution step is repeated iteratively until all agents satisfy $P_i(t) \le \rho_{eff}(t)$, ensuring that the hard-cap constraint is preserved in the final allocation.

Rather than adding complex rule-based constraints, the formulation ensures that the optimization itself converges to a stable operating point satisfying both efficiency and fairness—a computable fair division framework with fairness internalized within the optimization.

### 2.5. AHC++ (Adaptive Hard-Cap Controller++): An Adaptive Hard-Cap Controller for Dynamic Dominance Regulation

AHC++ (Adaptive Hard-Cap Controller++) is a dynamic control mechanism for probabilistically allocating limited computational resources (e.g., GPU time, tokens, bandwidth) among multiple agents or tenants. In this work, AHC++ first computes a Boltzmann–Softmax probability distribution from each agent's score or contribution input, and then updates the inverse temperature parameter β via feedback using the difference between the observed dominance signal and the target dominance in the system. In this process, β operates as a control variable that governs allocation concentration, and is dynamically adjusted to prevent the system's allocation structure from becoming excessively concentrated on any specific agent.

At each time step, the system computes Boltzmann–Softmax probabilities from observed scores, applies a hard-cap ensuring no agent exceeds the dominance upper bound, and renormalizes to sum to one, thereby preventing excessive concentration while maintaining system efficiency.

Simultaneously, the system updates β based on the error between observed and target dominance, forming a feedback control structure operating in the cycle: score input → probability computation → hard-cap → renormalization → β update.

The name reflects three features: **Adaptive**—β is dynamically adjusted based on the dominance error; **Hard-Cap**—direct probability limits suppress excessive concentration; **Controller**—the iterative feedback process guides the system toward a stable operating region.

Table 3 summarizes the correspondence between key terms and symbols used repeatedly in the description of AHC++.

**Table 3.** Correspondence between key terms and symbols used in AHC++.

| Category | Symbol | Economics | AI Engineering |
|---|---|---|---|
| Actor | $i$ | Individual | Agent / Tenant |
| Allocated object | $x_i(t)$ | Income / Wealth | Allocated resource amount<br>(e.g., GPU-time, tokens, bandwidth) |
| Determining factor | $z_i(t)$ | Capability / Contribution | Scheduling score |
| Control variable | $\beta$ | Distribution concentration | Inverse temperature<br>(softmax sharpness) |
| Policy weight | $\lambda$ | Redistribution intensity / Social weight | Fairness–efficiency trade-off weight |

***Note.*** The "++" denotes an extended control structure that incorporates safety mechanisms for stable operation in dynamic environments, including feasibility constraints (e.g., $\rho_{eff}$ ), probability clipping, renormalization, and smoothing.

### 2.5.1. Control Loop in Dynamic Environments (Illustrative Example)

At regular control intervals, AHC++ observes the system state, updates the control variable, and computes new allocations through the following stages.

**(i) Measurement**

At the beginning of each control interval, the system observes the top-1 dominance from the final allocation probabilities of the preceding interval—that is, after hard-cap application and renormalization. The dominance signal y(t) is defined as the largest value among these allocation probabilities:

$$y(t) = \max_i P_i(t) \tag{11}$$

In systems where probabilistic allocation is realized as actual selection outcomes, the dominance can also be estimated using the rolling realized share over the most recent $W$ intervals:

$$y_W(t) = \max_i \left( \frac{1}{W} \sum_{\tau=t-W+1}^{t} 1\{a(\tau) = i\} \right) \tag{12}$$

Here, $a(\tau)$ denotes the agent actually selected at time $\tau$, and $1\{\cdot\}$ is the indicator function. This expression represents the rolling realized dominance observed over the most recent W intervals. In the simulations presented in this work, y(t) is estimated using Eq. (12).

**(ii) Effective Target Dominance**

The effective target dominance $\rho_{eff}(t)$ is defined by considering the policy target $\rho_{target}(t)$ and the number of active agents $K(t)$ as follows:

$$\rho_{eff}(t) = \max\left( \rho_{target}(t), \frac{1}{K(t)} \right) \tag{13}$$

**(iii) Feedback Update**

The dominance error is defined as follows:

$$e(t) = y(t) - \rho_{eff}(t) \tag{14}$$

Here, y(t) is the top-1 dominance from the current allocation probabilities, and $\rho_{eff}(t)$ is the effective target dominance reflecting the policy target and system constraints.

The controller updates the inverse temperature parameter β, which governs allocation concentration, based on this error. The general form of the feedback update can be expressed as:

$$\beta(t+1) = Clip\big(\beta(t) - g(e(t))\big) \tag{15}$$

Here, $g(\cdot)$ is the control gain function, and the $Clip(\cdot)$ operation constrains β so that it does not exceed the allowable range $[\beta_{\min}, \beta_{\max}]$.

When a simple linear control gain is used, the above expression reduces to the following explicit control equation:

$$\beta(t+1) = Clip\big(\beta(t) - \eta\left[y(t) - \rho_{eff}(t)\right]\big) \tag{16}$$

Here, $y(t) = \max_i P_i(t)$ denotes the top-1 dominance from the current allocation probabilities, and $\rho_{eff}(t)$ denotes the target dominance reflecting the policy target and system constraints. The parameter $\eta$ is the control gain, which regulates the feedback intensity with respect to the dominance error.

This equation forms a negative feedback structure with respect to the dominance error. That is, when the dominance exceeds the target level (e(t) > 0), β decreases, relaxing the concentration of the probability distribution; conversely, when the dominance falls below the target, β increases, strengthening allocation concentration.

In practical implementations, additional stabilization techniques such as asymmetric gain, smoothing, or reference tracking may also be applied.

**(iv) Probability Computation and Resource Allocation (Allocation)**

Using the updated $\beta(t)$ and scores $z_i(t)$, the resource allocation is computed in the following steps.

First, the primary probabilities are computed using the Boltzmann–Softmax distribution:

$$P_i'(t) = \text{softmax}_i\left(\beta(t)\mathbf{z}(t)\right) = \frac{e^{\beta(t)\cdot z_i(t)}}{\sum_j e^{\beta(t)\cdot z_j(t)}} \tag{17}$$

A hard-cap constraint is then applied to ensure that no individual agent's share exceeds the upper bound:

$$\tilde{P}_i(t) = \min\left(P_i'(t), \rho_{eff}(t)\right) \tag{18}$$

Finally, renormalization is performed so that the total probability sums to one:

$$P_i(t) = \frac{\tilde{P}_i(t)}{\sum_j \tilde{P}_j(t)} \tag{19}$$

As in Section 2.4, redistribution is applied only to non-capped agents to ensure the hard-cap constraint is preserved after renormalization.

**Algorithm 1. AHC++ Control Procedure**

At each control interval, the following procedure is executed:

1. Observe the dominance $y(t) = \max_i P_i(t)$ from the current allocation probabilities.
2. Compute the effective target dominance $\rho_{eff}(t)$ using the policy target and the number of active agents.
3. Update the inverse temperature parameter $\beta(t)$ using the dominance error $e(t) = y(t) - \rho_{eff}(t)$.
4. Compute the softmax probabilities $P_i'(t)$ using the updated $\beta(t)$ and scores $z_i(t)$.
5. Apply the hard-cap $\rho_{eff}(t)$ to the computed probabilities.
6. Renormalize the probabilities after hard-cap application to obtain the final allocation probabilities $P_i(t)$.

The resulting $P_i(t)$ constitutes the resource allocation probabilities at the current time step. Through this control structure, AHC++ is able to form a stable allocation structure that suppresses dominance runaway of specific agents online, while preventing excessive degradation of performance metrics. From this perspective, AHC++ can be understood not as a simple probabilistic resource allocation algorithm, but as a feedback control mechanism that stabilizes allocation dynamics under dominance constraints.

# 3. Simulation and Results

This section examines, through numerical experiments, how the proposed Boltzmann–Softmax fair allocation mechanism regulates the trade-off between efficiency and fairness in AI resource allocation problems. All experiments were conducted in a Python environment, and the macroscopic dynamics of the system were analyzed from the perspective of agent-based modeling (ABM) in complexity economics.

### 3.1. Experimental Configuration

This subsection summarizes the configuration and parameter settings of the numerical experiments. The experiments model a population of agents (or tenants) competing for limited computational resources (GPU time, memory, bandwidth, etc.). The population size is set to $N = 1000$, and reflecting the heavy-tailed and rank-based characteristics observed in real-world AI ecosystems, the contribution score (or logit score) $z_i$ is defined as a logarithmic decreasing function of rank $i \in \{1, \cdots, N\}$:

$$z_i = -\ln(i) \quad \text{(where } i \in \{1, \ldots, N\}\text{)} \tag{20}$$

In this simulation, both the static and dynamic models use the same number of agents, $N = 1000$. The static model computes allocation statistics for N agents, and the dynamic model likewise simulates an environment in which the same $N = 1000$ tenants compete for resources.

Substituting the above contribution score definition into the Boltzmann–Softmax allocation formula, a power-law allocation naturally emerges without additional scaling. As an illustrative example, $P_i(\beta)$ can be expressed as follows:

$$P_i(\beta) = \frac{e^{\beta \cdot z_i}}{\sum_{j=1}^{N} e^{\beta \cdot z_j}} = \frac{i^{-\beta}}{\sum_{j=1}^{N} j^{-\beta}} \tag{21}$$

That is, the inverse temperature β is not merely a tuning parameter but operates as a core control variable that directly determines the slope of the resource allocation curve. By adjusting β, the structural concentration of the system—that is, the level of inequality or dominance—can be continuously controlled.

To jointly consider efficiency and fairness, this work defines a unified loss function (or unified objective function) and conducts evaluation and comparison under different policy weights $\lambda$. For example, the unified loss can be defined as follows:

$$L_{tot}(\beta, \lambda) = \lambda \cdot L_{eff}(\beta) + (1-\lambda) \cdot L_{ineq}(\beta), \quad \lambda \in [0,1] \tag{22}$$

The efficiency loss $L_{eff}$ and the fairness loss $L_{ineq}$ are constructed from the efficiency metric $Eff(\beta)$ and the fairness (entropy) metric $\mathrm{Eq}(\beta)$ in Table 4. Here, the efficiency loss $L_{eff}$ is a term that increases when resources are not sufficiently allocated to high-contribution (high $z_i$) agents, and the fairness loss $L_{ineq}$ is an entropy-based loss term that increases when resource concentration is not adequately suppressed. The policy weight $\lambda$ serves as a parameter that adjusts the relative importance of these two terms.

Table 4 summarizes the scenario specifications and key parameters commonly used across the simulations in Section 3.

**Table 4.** Simulation settings and workload parameters.

| Component | Setting |
|---|---|
| **Macro allocation curves (Figures 2–6)** | Population size N = 1000; rank-based logit $z_i = -\ln(i)$; β scan: linspace(0.05, 10.0, 400). |
| **Efficiency metric** | $\mathrm{Eff}(\beta) = \sum_{i=1}^{N_{pop}} P_i(\beta)\cdot \tilde{z}_i$, where $\tilde{z}_i = \frac{z_i - z_{\min}}{z_{\max} - z_{\min}} \in [0,1]$. |
| **Fairness metric (entropy-based equality)** | $\mathrm{Eq}(\beta) = \frac{H(P(\beta))}{\log(N_{pop})}$, where $H(P(\beta)) = -\sum_{i=1}^{N_{pop}} P_i(\beta)\log P_i(\beta)$. |
| **Queueing workload (Figures 7–11)** | Number of tenants $N = 1000$; simulation length T = 4200; service window W = 400; stable interval $t \ge 1600$. |
| **Arrivals** | Poisson arrivals: base arrival rate = 3.0 (jobs/step); tenant selection via Zipf popularity (α = 1.10). Arrival rate amplified ×1.80 during burst interval (t = 1000–1500). |
| **Service times** | Log-normal service: mean 2–12 (rank-dependent), jitter σ = 0.35. |
| **Contribution** | Fixed scores $z_i = -\ln(i)$; during abuse interval (t = 2200–2600), top tenant scoreBoost = +6.0, forceProb = 0.55. |
| **Scenario events** | burst (t = 1000–1500, ×1.80), policy change (t = 1400, λ: 0.50 → 0.75), abuse (t = 2200–2600), capacity reduction (t = 2800–3200, ×0.75). |
| **Dominance constraint** | Top-1 share measured over rolling window W, with $\rho_{target}$ determined by λ, ranging over [0.12, 0.40]. |
| **AHC++ control parameters** | $\beta_{min}$ = 0.1, $\beta_{max}$ = 8.0, control gain η = 6.0, smoothing coefficient = 0.85, reference tracking coefficient = 0.15. |
| **Random seeds** | Main benchmark: 10 seeds; sensitivity analysis: 8 seeds; policy comparisons conducted on identical workloads per seed. |

### 3.2 Results: Separation of Static and Dynamic Models and Analytical Flow

The policy target dominance is set as $\rho_{\text{target}}(t)$, and the effective target dominance is defined as $\rho_{eff}(t) = \max\left(\rho_{\text{target}}(t), \frac{1}{K(t)}\right)$. The "selected operating point" referred to in the text denotes the $(\lambda, \beta)$ combination chosen according to the criterion of (i) satisfying the dominance constraint $y(t) \le \rho_{eff}(t)$, and (ii) maximizing the efficiency metric $\mathrm{Eff}(\beta)$ or minimizing the unified loss $L_{\text{total}}(\lambda, \beta)$.

The results are organized into static and dynamic models. The static model constructs the “loss landscape” in a setting stripped of time, queues, and disturbances, characterizing where the efficiency–fairness trade-off arises and identifying the stability corridor—a region where solutions with nearly identical total loss form a band rather than a single optimum. The dynamic model then verifies whether these structures hold under realistic disturbances (arrival rate fluctuations, policy changes, adversarial traffic, capacity degradation), examining the trajectory along which AHC++ adjusts parameters to maintain target fairness, or the conditions under which limitations arise. If the static model serves as an "explainable map," the dynamic model is the "driving log" of a system navigating that map.

In the static model, the inverse temperature $\beta$ governs allocation concentration (smaller $\beta \rightarrow$ flatter, more equal distribution; larger $\beta \rightarrow$ winner-takes-all), while $\lambda$ sets the relative emphasis on efficiency versus fairness. The dynamic results evaluate not $\beta(t)$ itself, but how well the observed top-1 dominance satisfies the target constraint, and to what extent throughput, backlog, and latency are maintained. Constraint violations are also summarized as AUC for quantitative policy comparison.

The connection between static and dynamic analyses provides interpretable calibration: the optimal path and stability corridor serve as baselines for interpreting which $\beta$ region the dynamic operating point corresponds to, and whether the environment is pushing the system outside the corridor. Results are presented in the following order: the static model establishes the loss structure and stable operating range; the dynamic model compares policy performance under disturbances; finally, both are synthesized to verify structural consistency.

### 3.2.1 Static Model Results (Structure Only)

Figure 2 shows how $\beta$ governs the shape of the allocation probability distribution. At low $\beta$, curves are nearly horizontal (high-entropy, near-uniform). At high $\beta$, probability concentrates sharply on top ranks (low-entropy, winner-takes-all). Intermediate $\beta$ produces a compromise with gradual tail decay resembling the Zipf reference. This confirms $\beta$ as a core control knob regulating entropy and concentration.

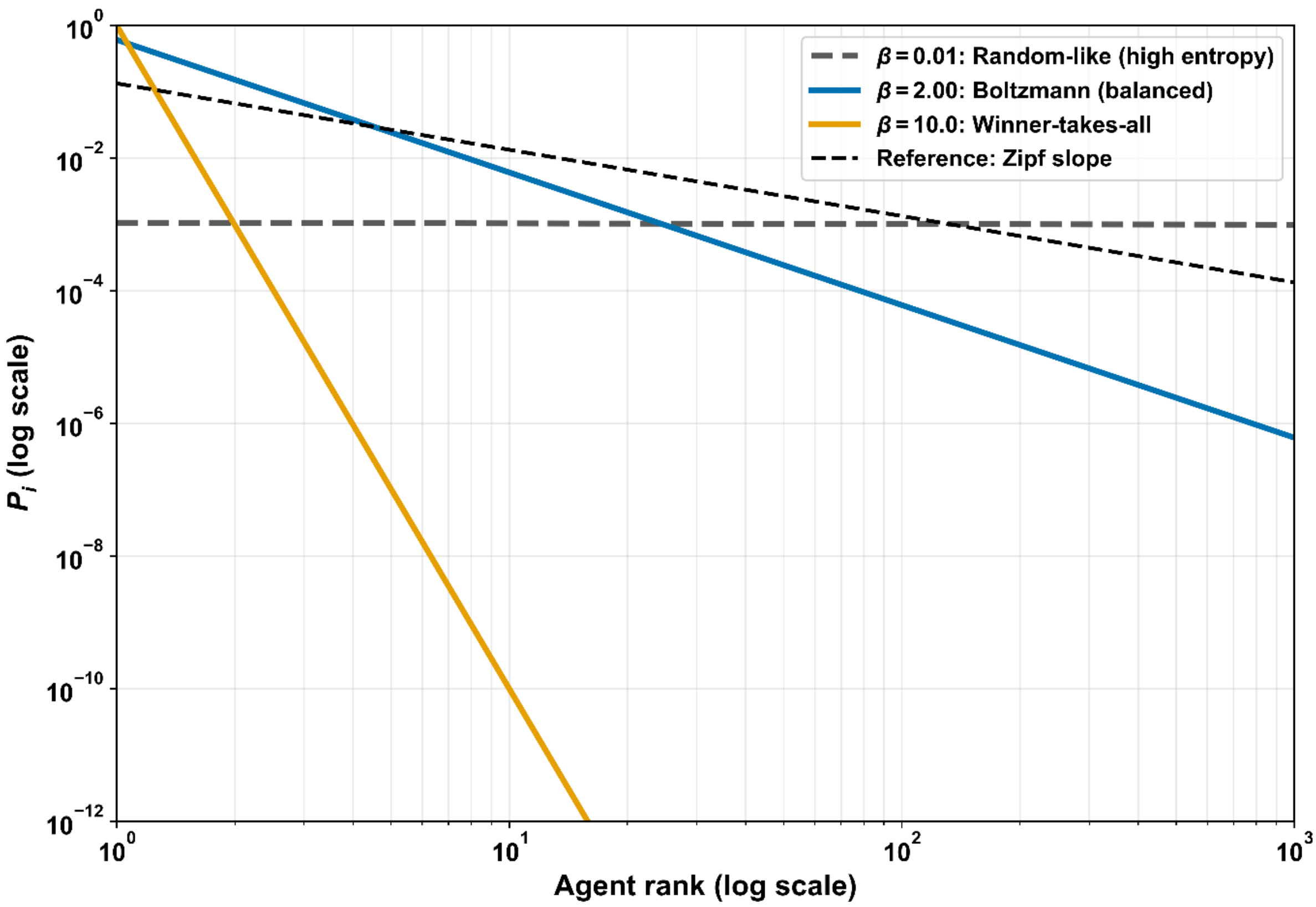

**Figure 2.** Variation of Boltzmann–Softmax allocation probabilities with inverse temperature β. Concentration is continuously regulated by β; the dashed line is the Zipf-slope reference.

Figure 3 plots efficiency loss, fairness loss, and λ-weighted total loss as functions of β. The two component losses move in opposite directions, confirming the trade-off. For each λ, a distinct β minimizes total loss, shifting rightward as λ increases. Near-optimal bands around each minimum suggest structural buffering—the basis for the stability corridor.

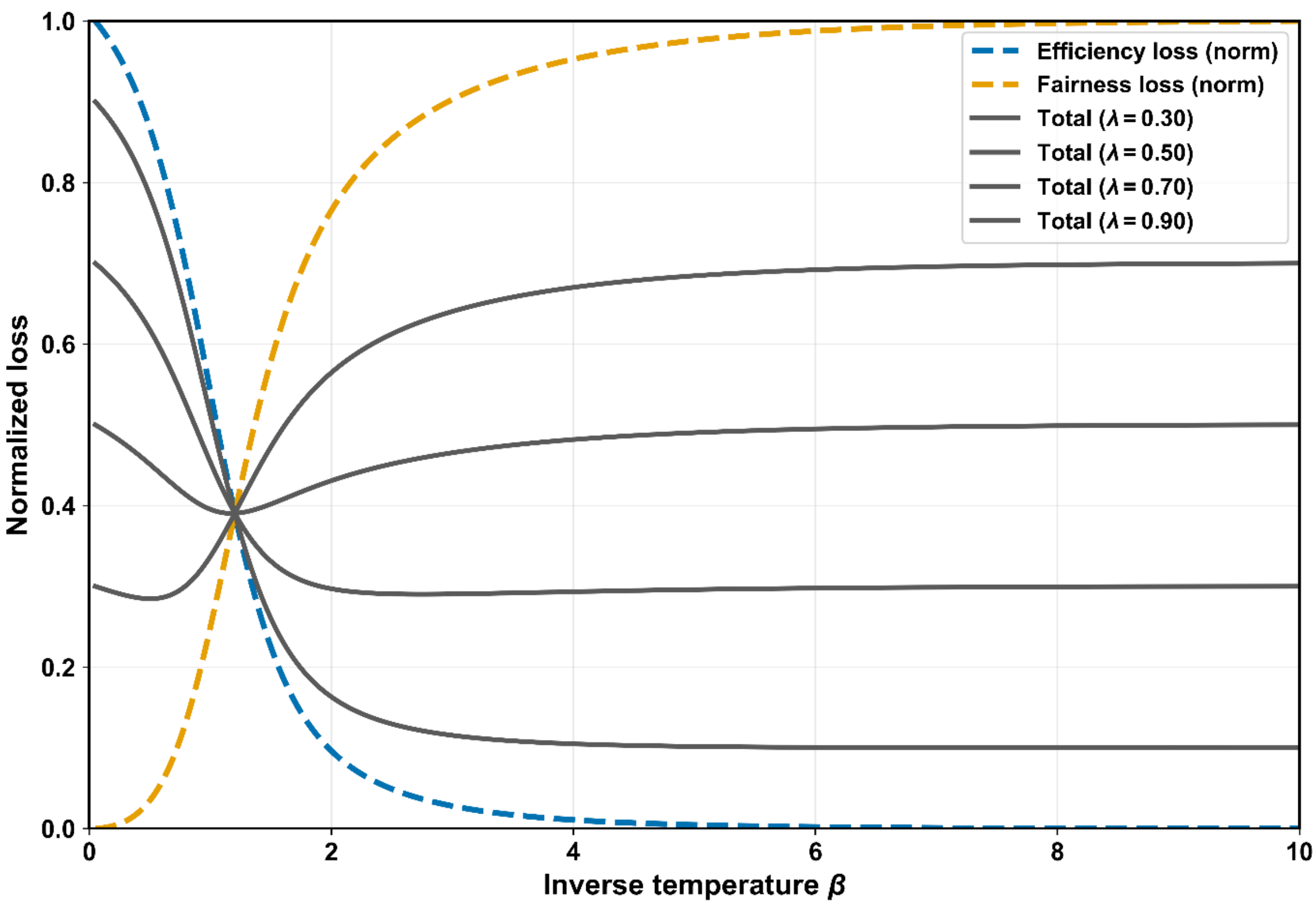

**Figure 3.** Static loss components and λ-weighted total loss vs. β. The minimum-loss β shifts with policy weight λ.

Figure 4 displays total loss as a heatmap in the (β, λ) plane. Excessively small β at high λ increases efficiency loss; excessively large β at low λ increases fairness loss. The static optimal path β*(λ) (black dashed) increases monotonically with λ. The light-blue shaded region between yellow boundaries defines the Stability Corridor—a connected low-loss zone where β can be adjusted without sharp performance degradation.

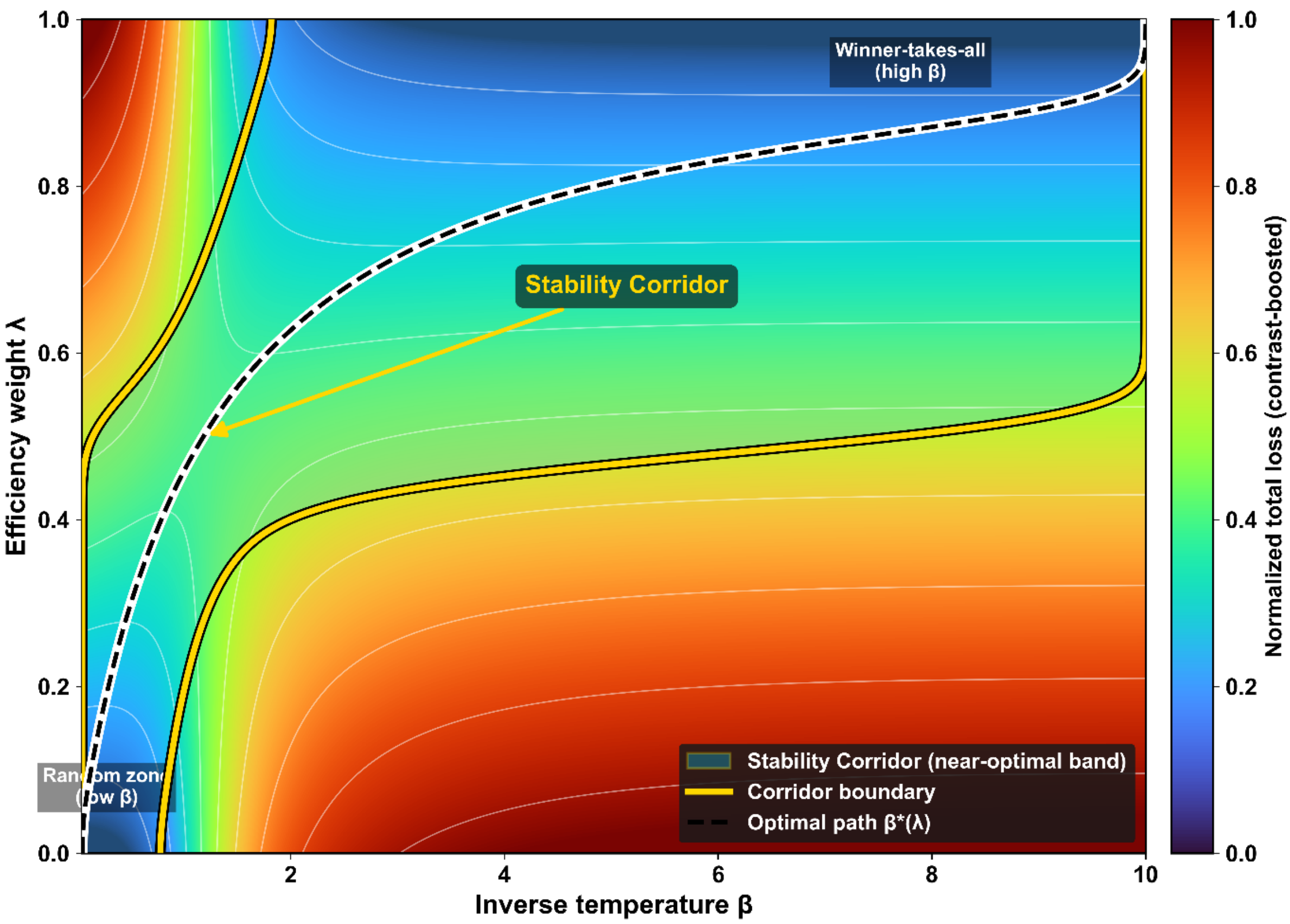

**Figure 4.** Total loss heatmap and Stability Corridor in the (β, λ) plane. Black dashed: optimal path β*(λ). Yellow lines: corridor boundaries.

Figure 5 bridges static analysis and dynamic control. Panel (a) shows the Pareto frontier with a near-optimal cloud—operating points of nearly identical total loss forming a band, the static basis for the stability corridor. The star marks $\beta^*(\lambda = 0.60)$, the dynamic control reference. Panel (b) contrasts post-shock trajectories: under fixed policy, the system may drift to extreme concentration and collapse; under AHC++, it returns to the stability zone.

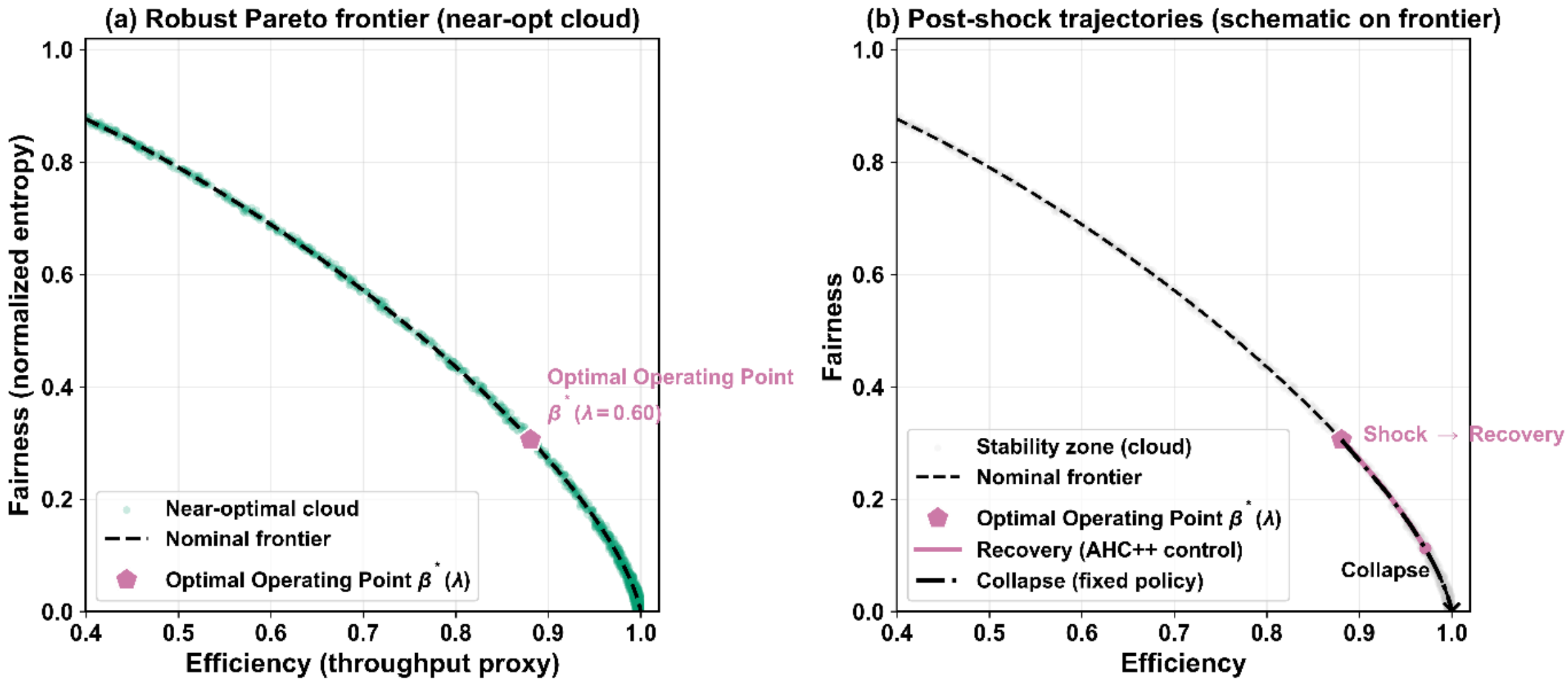

**Figure 5.** Pareto frontier, near-optimal cloud, and post-shock trajectories. (a) ★ = β*(λ = 0.60). (b) AHC++ recovers to stability zone; fixed policy collapses (×).

Figure 6 shows the efficiency–fairness relationship as β varies. The horizontal axis is fairness (normalized entropy); the vertical axis is efficiency (normalized weighted sum). As β varies, attainable states form a continuous trade-off frontier. At small β, the distribution is uniform: high fairness, lower efficiency. As β increases, probability concentrates on top agents: higher efficiency, reduced fairness. The dashed top-1 dominance line shows maximum allocation share increasing rapidly with β. This confirms β as a control variable governing concentration and fairness, providing the theoretical basis for its use in dynamic control.

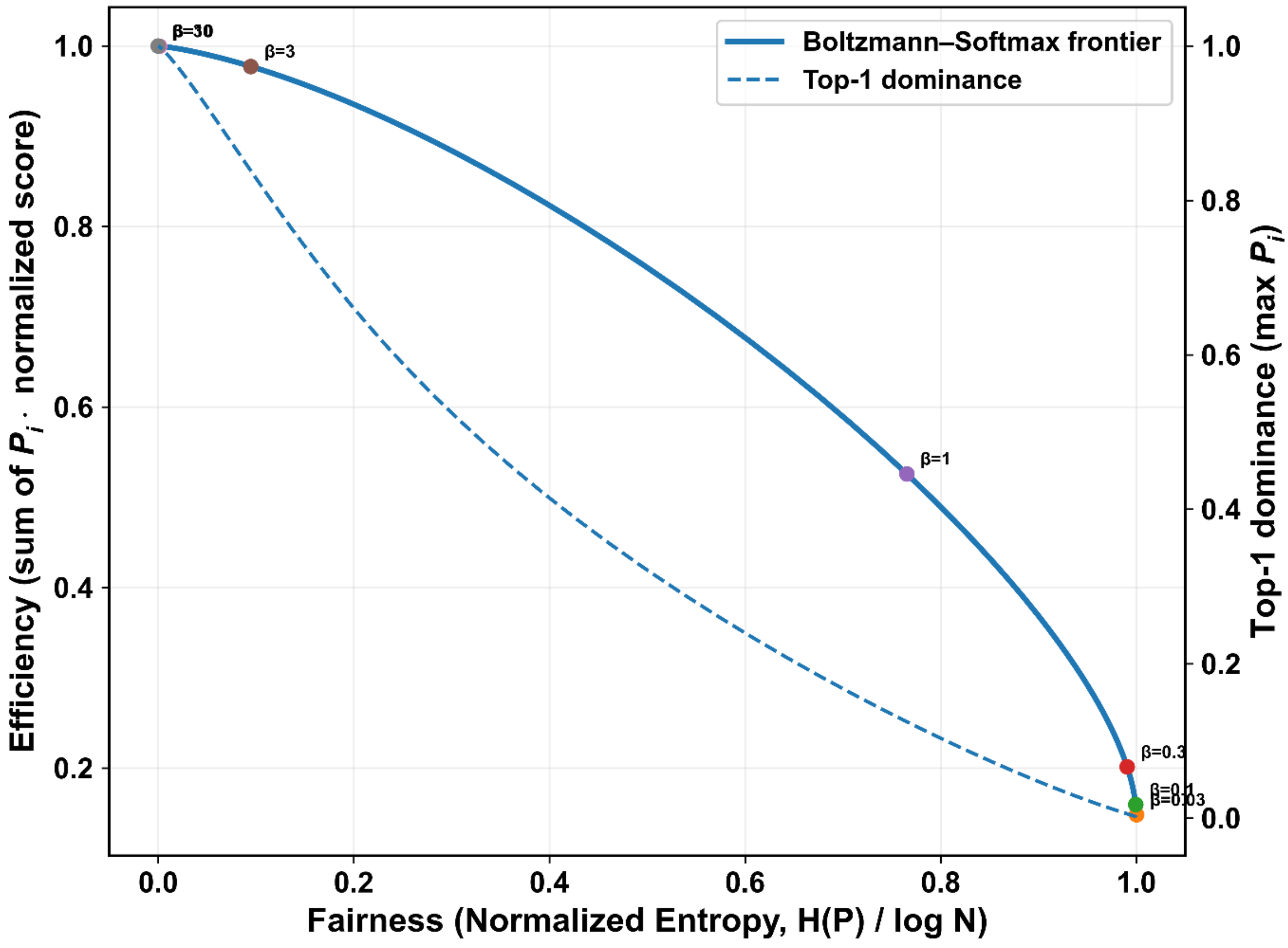

**Figure 6.** Efficiency–fairness frontier generated by the Boltzmann–Softmax rule. As β varies, attainable states trace a continuous trade-off curve; the dashed line shows top-1 dominance.

### 3.2.2 Dynamic Model Results (Structure Only)

Figure 7 compares the dominance control performance of four policies in a time-varying environment. RoundRobin allocates resources in a cyclic manner, aiming for uniform service on average but unable to directly incorporate score information or respond to disturbances. Greedy(z) prioritizes resource assignment to the tenant with the highest current score, which can be advantageous for throughput but readily leads to concentration of occupancy toward specific agents. FixedBetaSoftmax incorporates score information but keeps β fixed, limiting its adaptability to environmental changes. In contrast, AHC++ jointly adjusts β and the hard-cap based on observed dominance, actively tracking the target upper bound. Across event intervals, Greedy(z) and FixedBetaSoftmax maintain relatively high dominance due to score-based preferences and fixed β. During the abuse interval, both exhibit sharp target-exceeding spikes, whereas AHC++ suppresses violations through hard-cap application and error-based β updates. After capacity reduction, AHC++ reconverges below the target.

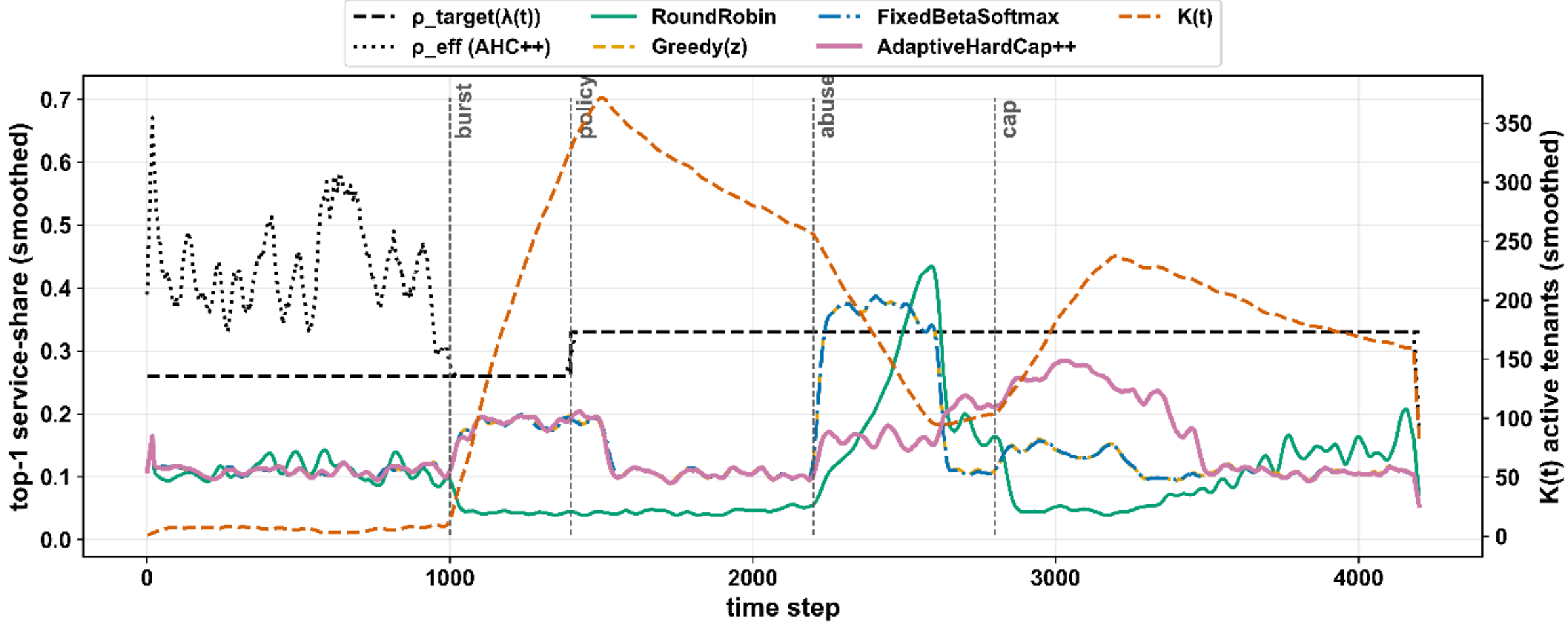

**Figure 7.** Smoothed top-1 dominance under time-varying events. Bold dashed: policy target. Dotted: effective target (AHC++). Red dashed (right axis): K(t).

Figure 8 directly compares FixedBetaSoftmax and AHC++. Both start with low dominance, but after burst and policy change, the fixed-β policy cannot adapt. During abuse, FixedBetaSoftmax sharply exceeds the target; AHC++ suppresses both magnitude and duration of violations. After capacity reduction, AHC++ readjusts and reconverges below the target.

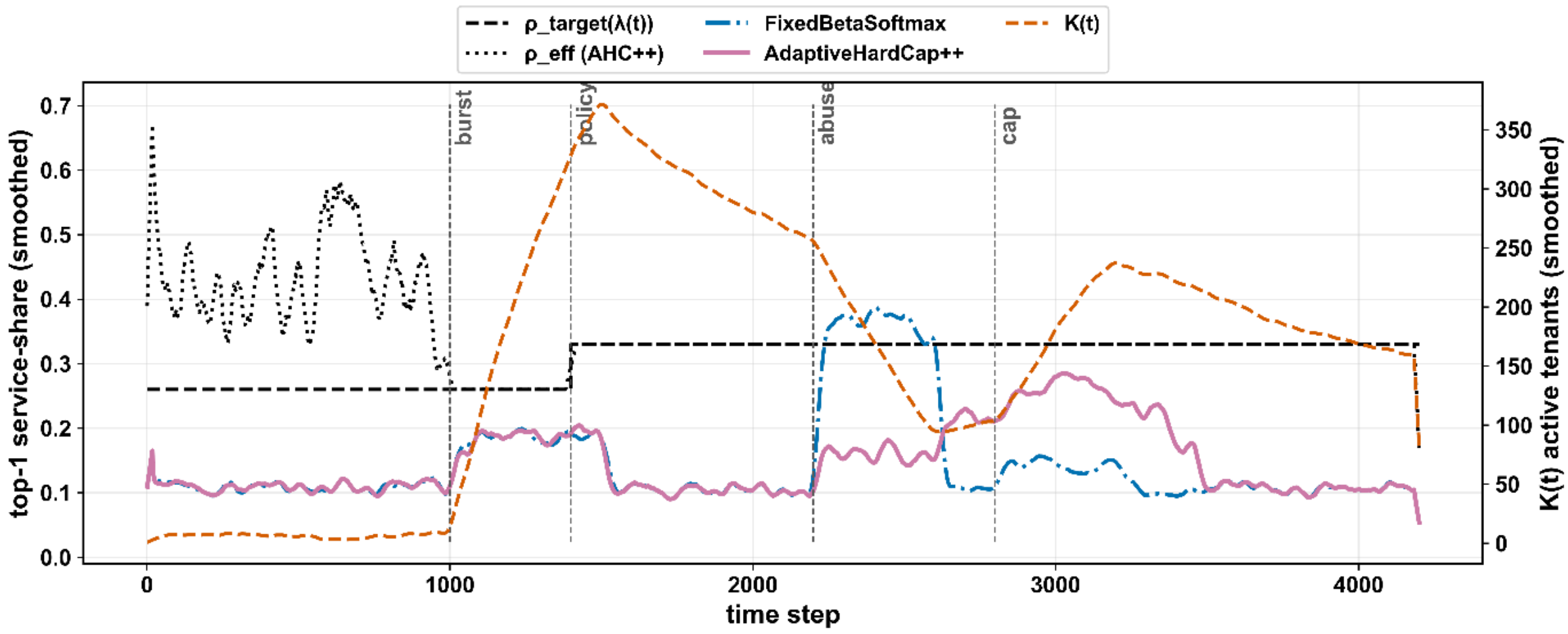

**Figure 8.** FixedBetaSoftmax vs. AHC++ dominance comparison over time.

Figure 9 shows AHC++'s control input $\beta(t)$ alongside the time-varying policy weight $\lambda(t)$. $\beta(t)$ drops at event boundaries (temporarily relaxing concentration) and recovers gradually. When λ shifts toward efficiency, $\beta(t)$ rises accordingly. Extended high-β intervals reflect conditions demanding greater concentration.

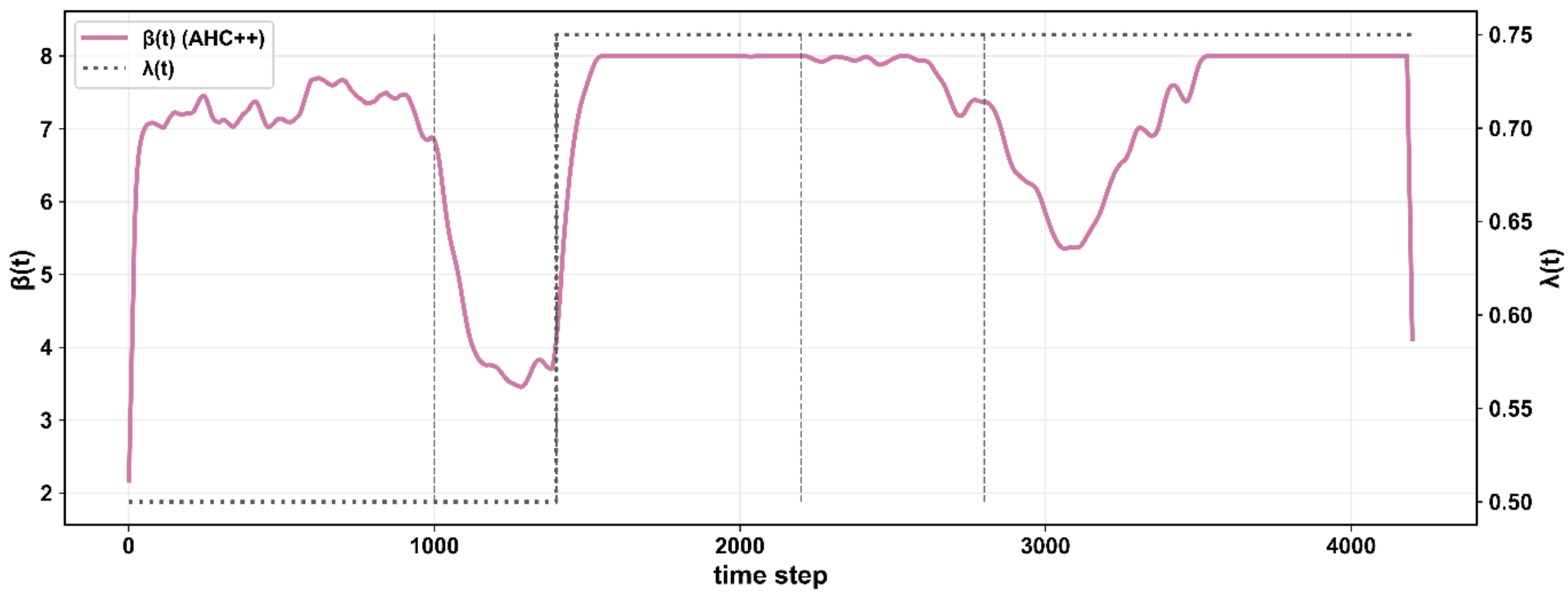

**Figure 9.** β(t) trajectory of AHC++ under time-varying λ(t). Vertical lines: event onsets.

Figure 10 compares how the number of unprocessed jobs in the system (backlog) accumulates and recovers across policies when time-varying events such as burst, policy change, abuse, and capacity reduction occur. After the burst onset ($t \approx 1000$), backlog increases sharply across all policies, reflecting the typical accumulation pattern that arises when the arrival rate momentarily exceeds processing capacity. After the burst ends ($t \approx 1500$), the backlog transitions into a declining phase, but the rate of decrease and the residual backlog level differ across policies. Greedy(z) tends to reduce the backlog relatively quickly after the burst, while FixedBetaSoftmax and AdaptiveHardCap++ exhibit intermediate recovery speeds. RoundRobin maintains a higher backlog level for a longer period in the same interval, and its recovery (drain) speed is observed to be relatively slower compared to score-based priority policies.

During abuse and capacity reduction events, backlog fluctuates again due to changes in system state. AHC++ shows temporary rises followed by decreases, consistent with dynamically adjusting the throughput–latency trade-off under dominance constraints.

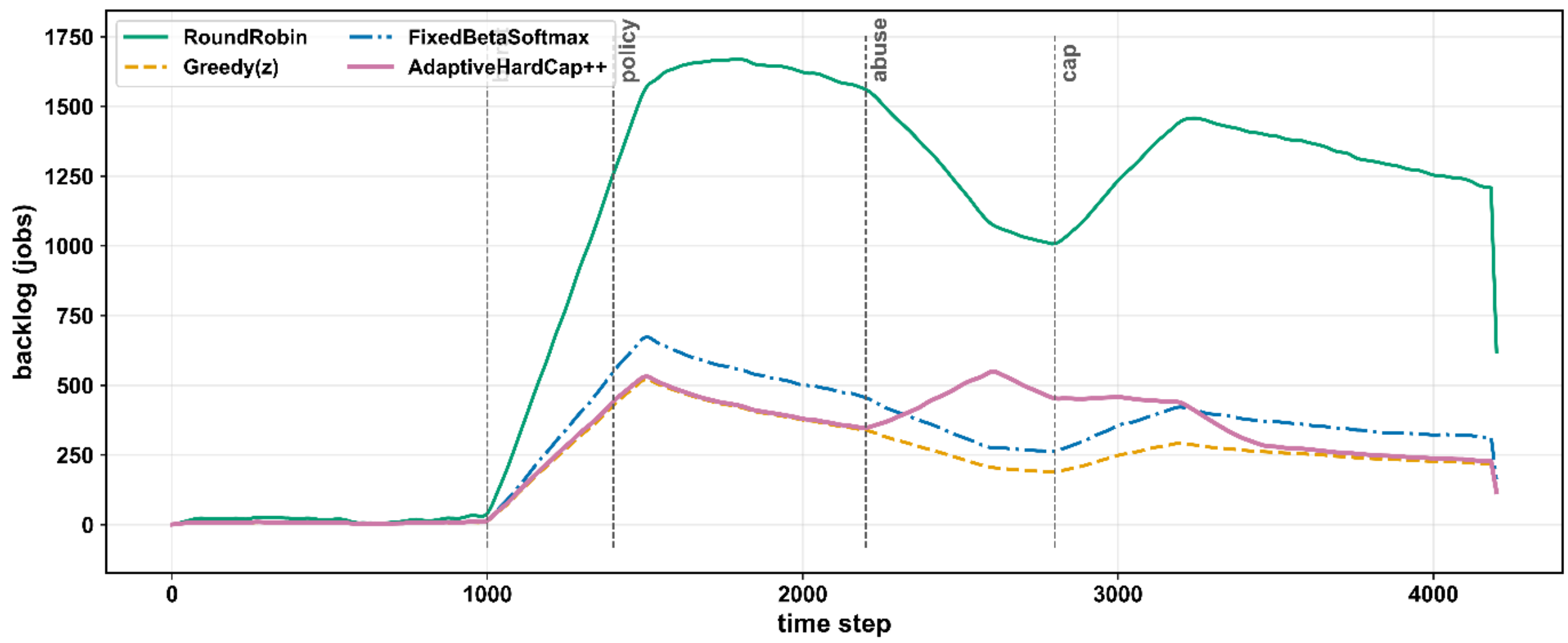

**Figure 10.** Backlog dynamics across policies under time-varying events.

Figure 11 presents a sensitivity analysis in which the burst factor (arrival spike multiplier) is varied and the cumulative constraint violation in the stable interval ($t \geq 1600$) is summarized as AUC. Here, $AUC_{\text{target}}$ denotes the cumulative exceedance obtained by integrating over time the positive deviation of the observed top-1 dominance above the policy target upper bound $\rho_{\text{target}}$; smaller values indicate more stable satisfaction of the target fairness constraint. $AUC_{eff}$ denotes the exceedance relative to the effective upper bound $\rho_{\text{eff}}$ that AHC++ applies internally, representing the degree of violation against a criterion adjusted to reflect "achievable fairness" based on factors such as the instantaneous number of active tenants $K(t)$.

The results show that FixedBetaSoftmax maintains roughly similar $AUC_{\text{target}}$ values even as the burst factor varies from 1.4 to 2.2, exhibiting a tendency to be structurally unable to mitigate dominance exceedance in response to changes in disturbance intensity. In contrast, AHC++ achieves $AUC_{\text{target}}$ and $AUC_{\text{eff}}$ values converging to nearly zero at burst factors of 1.8 and 2.2, demonstrating strong compliance performance in which the target upper bound is effectively satisfied continuously in the stable interval even in the presence of arrival rate surges.

Meanwhile, at a burst factor of 1.4, the $AUC_{\text{target}}$ value of AHC++ is observed to be relatively large, whereas $AUC_{\text{eff}}$ under the same condition is lower. This suggests that the target violations primarily occurred in intervals where the system internally follows the $\rho_{\text{eff}}$ criterion, but $\rho_{\text{target}}$ is set more stringently (e.g., situations where $K(t)$ is small and top-1 dominance is structurally bound to be high).

In summary, this sensitivity analysis demonstrates that (i) fixed-temperature policies have difficulty systematically reducing constraint violations even as disturbance intensity varies, and (ii) AHC++ actively satisfies the dominance constraint in response to environmental conditions, and is capable of nearly eliminating cumulative violations in the stable interval, particularly under moderate-to-high burst conditions.

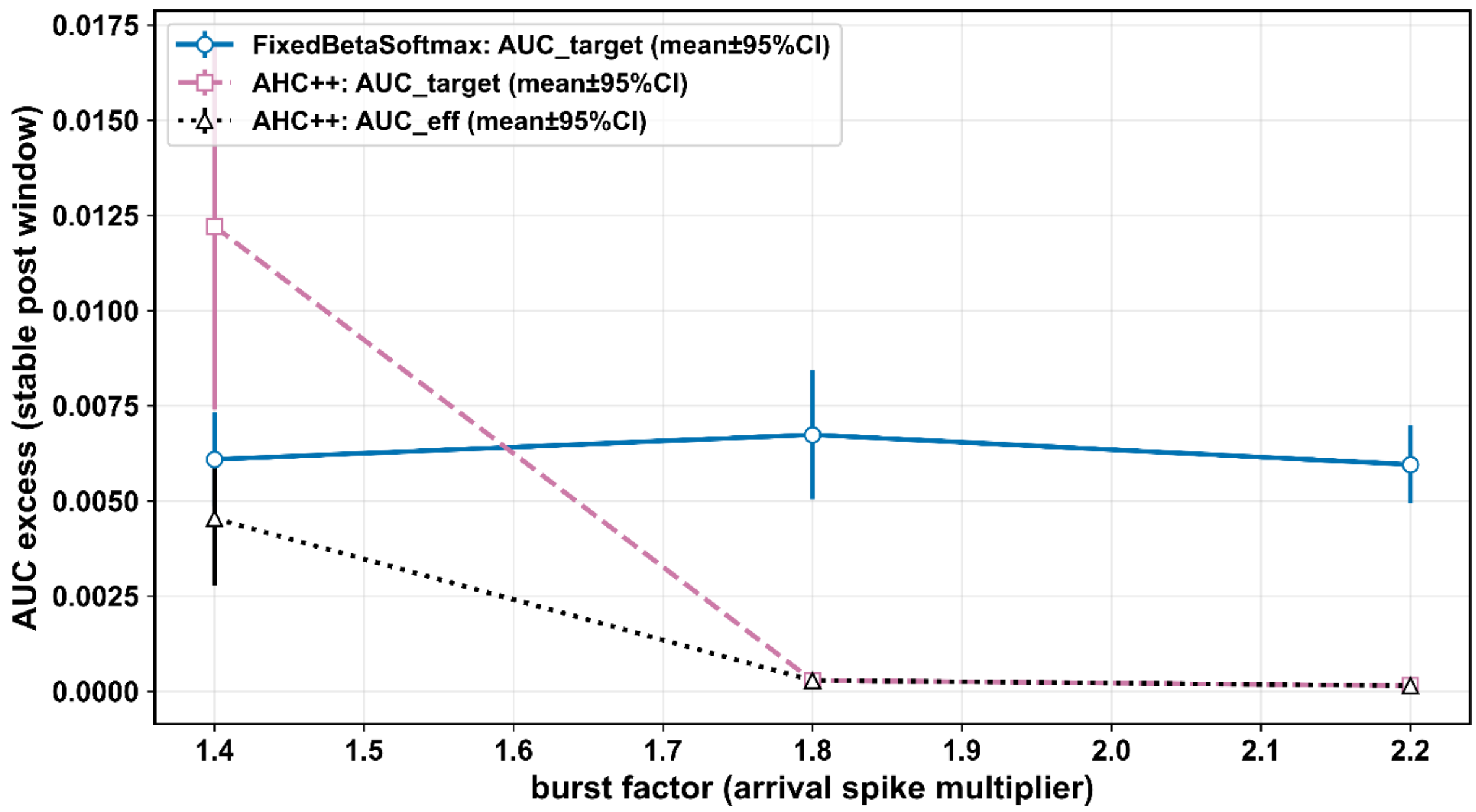

**Figure 11.** Cumulative dominance constraint exceedance (AUC) vs. burst factor in the stable interval. AHC++ nearly eliminates violations; FixedBetaSoftmax remains constant (error bars: mean ± 95% CI).

Taken together, the results from Figures 7–11 show that under dynamic disturbances, each policy exhibits a different balance between robustness of dominance constraint compliance and service quality (throughput and latency). To compare these more directly, the four policies were repeatedly simulated under identical settings, and the key performance and stability metrics are summarized in Table 5. The simulations were conducted with 1000 tenants over 4200 time steps, and dominance was measured using a rolling window of length 400. The policy weight λ(t) is switched during the simulation, and the corresponding target dominance upper bound is adjusted accordingly. Disturbances consist of arrival rate burst, abuse, and capacity reduction events, and result evaluation was performed in the stable interval to reduce transient effects. All values are reported as the mean and 95% confidence interval over 10 seed repetitions.

**Table 5.** Performance and stability metrics by policy (mean ± 95% confidence interval).

| Metric | RoundRobin | Greedy(z) | FixedBetaSoftmax | AdaptiveHardCap++ |
|---|---|---|---|---|
| **thrpt** | 2.9972 ± 0.0335 | 3.2325 ± 0.0164 | 3.2093 ± 0.0157 | 3.2304 ± 0.0165 |
| **meanLat** | 228.7 ± 12.1 | 41.2 ± 2.9 | 80.1 ± 4.5 | 65.7 ± 2.4 |
| **p95Lat** | 1249.5 ± 104.4 | 184.3 ± 35.7 | 482.4 ± 26.8 | 383.1 ± 15.2 |
| **maxTop1** | 0.534 ± 0.103 | 0.473 ± 0.017 | 0.471 ± 0.017 | 0.358 ± 0.021 |
| **frac>tgt** | 0.045 ± 0.019 | 0.109 ± 0.006 | 0.109 ± 0.007 | 0.013 ± 0.008 |
| **AUC_tgt** | 0.00518 ± 0.00458 | 0.00591 ± 0.00091 | 0.00588 ± 0.00090 | 0.00027 ± 0.00021 |
| **AUC_eff** | 0.00518 ± 0.00458 | 0.00591 ± 0.00091 | 0.00588 ± 0.00090 | 0.00027 ± 0.00021 |
| **backEnd** | 1205 ± 152 | 217 ± 27 | 314 ± 41 | 226 ± 28 |

***Note.*** *N = 1000, T = 4200, W = 400, 10 seeds. Post-stabilization window: t ≥ 1600.*

Table 5 summarizes four policies under identical dynamic environments. Key findings: AHC++ achieves the strongest dominance control (maxTop1 = 0.358, approximately 24% lower than fixed-parameter policies at ~0.47) while maintaining throughput (3.2304) nearly identical to Greedy(z) (3.2325) and reducing constraint violation frequency by approximately eightfold (frac>tgt = 1.3% vs. 10.9%). The trade-off cost is moderate latency increase: meanLat and p95Lat are approximately 1.6× and 2.1× those of Greedy(z), respectively. RoundRobin shows low violation rates but substantially worse throughput, latency, and backlog. AHC++ thus operates near a stable equilibrium between fairness and performance. In this scenario, AUC_tgt and AUC_eff are identical across all policies. For AHC++, this is because K(t) was sufficiently large throughout the stable interval that $\rho_{eff}(t) = \max\left(\rho_{target}(t), 1/K(t)\right)$ reduced to $\rho_{target}(t)$ at every time step. For the baseline policies (RoundRobin, Greedy(z), FixedBetaSoftmax), $\rho_{eff}$ is not applied internally; both AUC metrics are therefore computed against the same ρ_target baseline. The two metrics would diverge in scenarios where K(t) is small enough that $1/K(t) > \rho_{target}(t)$, as illustrated in the burst factor sensitivity analysis (Figure 11, burst factor = 1.4).

### 3.2.3 Computational Scalability

This subsection evaluates whether the proposed mechanism possesses the computational efficiency required for repeated execution in large-scale AI systems, where the cost per allocation step is critical to practical viability. In this regard, the present analysis constitutes a computational scalability assessment designed to evaluate whether the proposed mechanism is not only "fair" but also "computable and scalable" in real operational environments.

Execution time was measured for a single allocation step—(i) Boltzmann–Softmax probability computation and (ii) hard-cap redistribution—as N increased from $10^2$ to $10^4$. Warm-up runs preceded repeated executions; median and IQR were computed. A pairwise dummy served as an $O(N^2)$ reference. All measurements used a single-CPU Python environment (Intel i7-8565U, 16 GB RAM, Windows 11).

Figure 12 shows that execution time is flat at small N (dominated by fixed overhead) and increases moderately thereafter. As N grew 100× ($10^2$ to $10^4$), execution time increased only ~5.5×—consistent with the vectorized probability-and-normalization structure. The pairwise dummy scales as $O(N^2)$, highlighting the structural advantage.

These results demonstrate that the proposed framework offers both practical computability and scalability, supporting its feasibility as a repeatedly executable allocation rule in large-scale multi-agent environments.

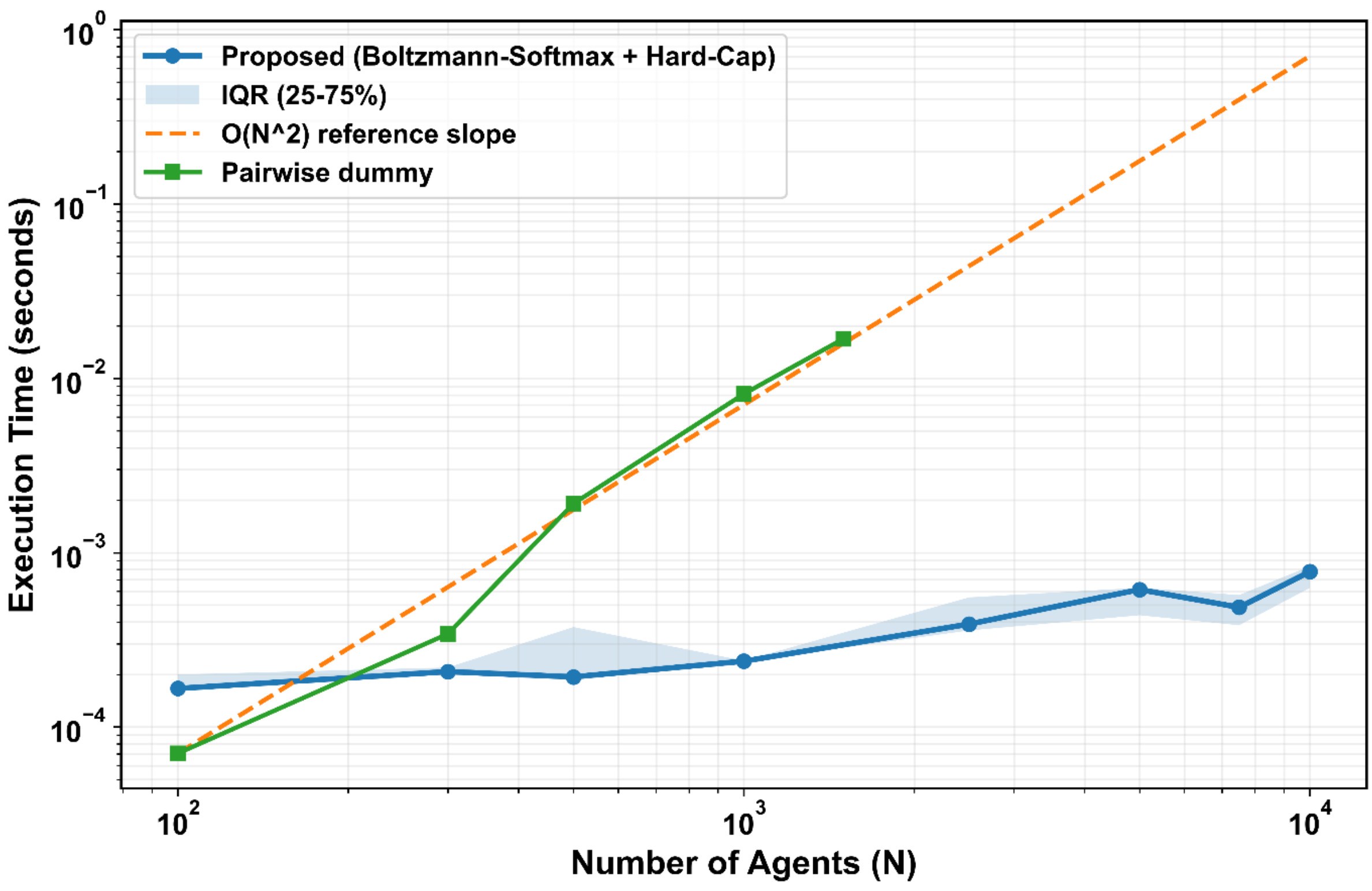

**Figure 12.** Computational scalability of a single allocation step ($N = 10^2$ to $10^4$). Proposed mechanism: ~5.5× for 100× agents. Pairwise dummy: $O(N^2)$.

# 4. Discussion

This work reformulated the resource allocation problem in AI systems from the perspective of statistical physics, and demonstrated through simulation that the Boltzmann–Softmax allocation rule combined with the AHC++ control algorithm can suppress dominance runaway while maintaining system performance. In particular, the loss landscape and near-optimal band derived from the static model were shown to be structurally connected to the control and recovery behavior of the dynamic model. These results suggest that the relationship between efficiency and fairness can be reframed as a problem of computable allocation rules and feedback control, rather than one of abstract ethical declarations or post-hoc regulation. The following subsections discuss the implications of these findings.

## 4.1 From Exogenous Regulation to Intrinsic Fairness

Existing AI fairness research has primarily relied on exogenous approaches that impose fairness from outside the system, such as post-processing of model outputs or constraint imposition during training (Zafar et al., 2017). Some studies have also presupposed efficiency and fairness as inherently conflicting objectives, designing systems in which a planner selects an appropriate compromise (Zheng et al., 2020). While such approaches can be effective under certain conditions, they require continuous maintenance of external constraints, and violations can easily arise when environmental changes or disturbances occur.

In contrast, this work internalizes fairness as a design variable by defining the allocation mechanism through the Boltzmann–Softmax rule. The inverse temperature parameter $\beta$ continuously regulates concentration and entropy, making fairness a tunable property within the allocation rule itself rather than a constraint appended after the fact.

The mathematical structure is consistent with the Boltzmann distribution and admits an information-theoretic interpretation through the maximum entropy principle—the least biased distribution under given constraints (Jaynes, 1957; Cover & Thomas, 2006). Fairness can thus be understood as a structural property embedded in the probabilistic allocation format itself, rather than the product of a particular heuristic choice.

The fairness objective is further operationalized through an observable metric—top-1 dominance—combined with online feedback control, transforming fairness from a declarative principle into a computable control target that the system can track and adjust in real time.

**4.2 Policy Implications and Robustness of the Stability Corridor**

In practice, policy targets are uncertain and shocks inevitable. If the optimum is overly parameter-sensitive, small perturbations could trigger monopolization. The static analysis reveals a stability corridor where total loss remains robust to changes in the efficiency weight, functioning as a buffer zone against policy errors.

From an operational perspective, real-world system administrators rarely know the exact optimal parameter setting; instead, they prefer a stable operating range within which safe adjustments can be made. The corridor provides precisely this: a practical guide for target setting and retuning, indicating within what range the policy weight $\lambda$ and the inverse temperature $\beta$ can be adjusted without incurring sharp performance degradation. After a disturbance, the corridor also indicates which $\beta$ region remains safe for recovery, offering interpretable guidance for post-shock parameter adjustment.

The corridor thus provides a quantitative basis for dynamic resilience, allowing the system to absorb parameter uncertainties without drifting toward monopolistic collapse. In this sense, the stability corridor extends beyond a mathematical concept and serves as a decision-support tool for AI governance—a quantitative map showing within what range it is safe to operate.

### 4.3 Computational Scalability and Real-Time Operational Feasibility

In large-scale AI systems, the computational cost of the allocation mechanism is critical for practical viability. The Shapley value, for instance, scales combinatorially with participant count, precluding real-time application. The Boltzmann–Softmax rule, structured around probability computation and normalization, maintains a simpler architecture.

In our single-CPU Python implementation, scalability analysis (Figure 12) shows that as agents increased 100× (from $10^2$ to $10^4$), execution time of a single allocation step grew by only ~5.5× (median over repeated runs), demonstrating moderate cost growth under the tested configuration.

The algorithm's parallel-friendly structure—based on vector operations and normalization rather than pairwise comparisons—indicates that efficient hardware utilization can be maintained even at larger scales, though performance on production-grade infrastructure remains to be verified.

Furthermore, the AHC++ controller dynamically adjusts the inverse temperature parameter and hard-cap based on observed dominance, making it applicable in real-time operational environments. These characteristics suggest that fairness constraints can be employed as operational metrics while maintaining system performance in GPU cluster or cloud infrastructure environments where thousands to tens of thousands of tenants compete. Accordingly, the framework constitutes a real-time fairness mechanism—computable and controllable—rather than a static allocation rule.

### 4.4 Limitations and Future Directions

This work has demonstrated that fairness can be treated as a computable and controllable operational variable by combining the Boltzmann–Softmax allocation rule with the AHC++ control structure. However, the results are primarily focused on confirming structural feasibility and operating principles, and are accompanied by several interpretive limitations.

First, the dynamic experiments remain stylized simulations that do not fully capture real GPU cluster or inference serving workloads. Future work should incorporate trace-driven workloads, tenant priorities, deadline sensitivity, heterogeneous task sizes, and straggler effects.

Second, scalability analysis covers single allocation steps, not end-to-end system latency. Operational verification through system integration or cluster-scale prototypes is needed.

Third, top-1 dominance alone does not capture all fairness dimensions. Complementary metrics (Jain's fairness index, starvation rate, tail latency, tenant-level consistency) should be jointly evaluated.

Fourth, the baseline policies (RoundRobin, Greedy(z), FixedBetaSoftmax) are simple policies that either do not explicitly consider fairness or use fixed parameters. This selection was intended to demonstrate the structural advantages of AHC++'s adaptive control relative to non-adaptive policies. However, the resource allocation literature has already proposed methodologies that explicitly consider fairness, such as proportional fairness, max-min fairness, and Dominant Resource Fairness (DRF).

Future research should more clearly delineate the relative advantages and scope of applicability of this framework through systematic comparison with such established baselines. In particular, analyzing under what conditions the Boltzmann–Softmax rule produces results that are similar to or complementary with existing fair allocation methodologies would contribute to clarifying the theoretical positioning of this framework.

Nevertheless, the significance of this work lies in formalizing fairness as a computational structure that can be adjusted within the allocation rule and feedback control. Future research can build upon this framework by connecting it to more realistic system environments and demonstrating practical competitiveness through comparison with diverse baseline policies.

# 5. Conclusion

This work proposed a computable fair division framework for large-scale AI resource allocation, combining the Boltzmann–Softmax allocation rule with the AHC++ feedback control structure. The inverse temperature parameter β serves as a computable control variable that dynamically governs the balance between efficiency and fairness.

Static analysis identified the stability corridor—a near-optimal operating region where performance remains robust to policy parameter variation—providing operators with a realistic and safe policy choice space. Dynamic simulations confirmed that AHC++ suppresses dominance concentration under disturbance environments while maintaining competitive throughput and backlog performance. Computational scalability analysis further demonstrated the feasibility of repeated execution in large-scale multi-agent environments.

End-to-end verification in real distributed infrastructure and generalization to diverse workloads remain as future tasks. Nevertheless, this work shows that the static loss landscape and dynamic control behavior can be interpreted within a unified analytical framework. Within this framework, computable fair division can function not merely as a normative ideal, but as a practical principle of operational design. In this regard, this work reframes the AI fairness discourse from an abstract normative concern into a concrete problem of computable system design and feedback control.

# References

Arrow, K. J. (1951). Social choice and individual values. Yale University Press.

Atkinson, A. B. (1970). On the measurement of inequality. Journal of Economic Theory, 2(3), 244–263.

Barocas, S., Hardt, M., & Narayanan, A. (2023). Fairness and machine learning: Limitations and opportunities. MIT Press.

Bommasani, R., Hudson, D. A., Adeli, E., Altman, R., Arora, S., von Arx, S., Bernstein, M. S., Bohg, J., Bosselut, A., Brunskill, E., Brynjolfsson, E., Buch, S., Card, D., Castellon, R., Chatterji, N., Chen, A., Creel, K. A., Davis, J. Q., Demszky, D., ... Liang, P. (2021). On the opportunities and risks of foundation models. arXiv. https://arxiv.org/abs/2108.07258

Cover, T. M., & Thomas, J. A. (2006). Elements of information theory (2nd ed.). Wiley.

Dwork, C., Hardt, M., Pitassi, T., Reingold, O., & Zemel, R. (2012). Fairness through awareness. In Proceedings of the 3rd Innovations in Theoretical Computer Science Conference (pp. 214–226).

Ensign, D., Friedler, S. A., Neville, S., Scheidegger, C., & Venkatasubramanian, S. (2018). Runaway feedback loops in predictive policing. In Proceedings of the 1st Conference on Fairness, Accountability and Transparency (pp. 160–171).

Feldman, M., Friedler, S. A., Moeller, J., Scheidegger, C., & Venkatasubramanian, S. (2015). Certifying and removing disparate impact. In Proceedings of the 21st ACM SIGKDD International Conference on Knowledge Discovery and Data Mining (pp. 259–268).

Ghorbani, A., & Zou, J. (2019). Data Shapley: Equitable valuation of data for machine learning. In K. Chaudhuri & M. Sugiyama (Eds.), Proceedings of the 36th International Conference on Machine Learning (Proceedings of Machine Learning Research, Vol. 97, pp. 2242–2251).

Hardt, M., Price, E., & Srebro, N. (2016). Equality of opportunity in supervised learning. In D. D. Lee, M. Sugiyama, U. V. Luxburg, I. Guyon, & R. Garnett (Eds.), Advances in neural information processing systems (Vol. 29, pp. 3315–3323).

Jaynes, E. T. (1957). Information theory and statistical mechanics. Physical Review, 106(4), 620–630.

Kleinberg, J., Mullainathan, S., & Raghavan, M. (2017). Inherent trade-offs in the fair determination of risk scores. In Proceedings of the 8th Innovations in Theoretical Computer Science Conference (ITCS 2017).

O'Neil, C. (2016). Weapons of math destruction: How big data increases inequality and threatens democracy. Crown.

Okun, A. M. (1975). Equality and efficiency: The big tradeoff. Brookings Institution Press.

OpenAI. (2023). GPT-4 technical report. arXiv. https://arxiv.org/abs/2303.08774

Park, J.-W. (2024). From physics to environmental policy: Exploring Boltzmann distribution for carbon trading permit allocation. Nakhara: Journal of Environmental Design and Planning, 23(1), Article 405.

Park, J.-W., & Kim, C. U. (2021). Getting to a feasible income equality. PLOS ONE, 16(3), e0249204.

Park, J.-W., Kim, C. U., & Isard, W. (2012). Permit allocation in emissions trading using the Boltzmann distribution. Physica A: Statistical Mechanics and Its Applications, 391(20), 4883–4890.

Park, J.-W., Kim, J. U., Ghim, C.-M., & Kim, C. U. (2022). The Boltzmann fair division for distributive justice. Scientific Reports, 12, 16179.

Samuelson, P. A. (1947). Foundations of economic analysis. Harvard University Press.

Shapley, L. S. (1953). A value for n-person games. In H. W. Kuhn & A. W. Tucker (Eds.), Contributions to the theory of games II (pp. 307–317). Princeton University Press.

Sitthiyot, T., & Holasut, K. (2025). A cross-country analysis of feasible income equality using the sigmoid function and the Boltzmann distribution. *PLOS ONE*, 20(8), e0329633.

Sutton, R. S., & Barto, A. G. (2018). Reinforcement learning: An introduction (2nd ed.). MIT Press.

Zafar, M. B., Valera, I., Gomez Rodriguez, M., & Gummadi, K. P. (2017). Fairness constraints: Mechanisms for fair classification. In A. Singh & J. Zhu (Eds.), Proceedings of the 20th

International Conference on Artificial Intelligence and Statistics (Proceedings of Machine Learning Research, Vol. 54, pp. 962–970).

Zheng, S., Trott, A., Srinivasa, S., Naik, N., Gruesbeck, M., Parkes, D. C., & Socher, R. (2020). The AI economist: Improving equality and productivity with AI-driven tax policies. arXiv. https://arxiv.org/abs/2004.13332